\documentclass{emulateapj}
\usepackage{graphics}
\usepackage{psfig}
\usepackage{ulem}
\usepackage{longtable}

\def\sn{{SN~1979C}}
\def\E{{\sl Einstein}}
\def\R{{\sl ROSAT}}
\def\A{{\sl ASCA}}
\def\C{{\sl Chandra}}

\def\X{{\sl XMM-Newton}}
\def\H{{\sl HST}}
\def\V{{\sl VLA}}
\def\VLBI{{\sl VLBI}}

\def\gs{\mathrel{\mathchoice {\vcenter{\offinterlineskip\halign{\hfil
$\displaystyle##$\hfil\cr>\cr\sim\cr}}}
{\vcenter{\offinterlineskip\halign{\hfil$\textstyle##$\hfil\cr
>\cr\sim\cr}}}
{\vcenter{\offinterlineskip\halign{\hfil$\scriptstyle##$\hfil\cr
>\cr\sim\cr}}}
{\vcenter{\offinterlineskip\halign{\hfil$\scriptscriptstyle##$\hfil\cr
>\cr\sim\cr}}}}}
\def\ls{\mathrel{\mathchoice {\vcenter{\offinterlineskip\halign{\hfil
$\displaystyle##$\hfil\cr<\cr\sim\cr}}}
{\vcenter{\offinterlineskip\halign{\hfil$\textstyle##$\hfil\cr
<\cr\sim\cr}}}
{\vcenter{\offinterlineskip\halign{\hfil$\scriptstyle##$\hfil\cr
<\cr\sim\cr}}}
{\vcenter{\offinterlineskip\halign{\hfil$\scriptscriptstyle##$\hfil\cr
<\cr\sim\cr}}}}}

\begin{document}

\title{Late-time X-Ray, UV and Optical Monitoring of Supernova 1979C}

\author{Stefan~Immler\altaffilmark{1},
Robert A. Fesen\altaffilmark{2},
Schuyler D. Van Dyk\altaffilmark{3},
Kurt W. Weiler\altaffilmark{4}, 
Robert Petre\altaffilmark{1},
Walter H. G. Lewin\altaffilmark{5},
David Pooley\altaffilmark{6},
Wolfgang Pietsch\altaffilmark{7},
Bernd Aschenbach\altaffilmark{7},
Molly C. Hammell\altaffilmark{2} and
Gwen C. Rudie\altaffilmark{2}
}

\altaffiltext{1}{Exploration of the Universe Division, 
X-Ray Astrophysics Laboratory, Code 662, 
NASA Goddard Space Flight Center, Greenbelt, MD 20771, USA}
\email{immler@milkyway.gsfc.nasa.gov}
\altaffiltext{2}{Department of Physics and Astronomy,
6127 Wilder Laboratory, Dartmouth College, Hanover, NH 03755, USA}
\altaffiltext{3}{Spitzer Science Center/Caltech, 220-6, Pasadena, CA 91125, USA}
\altaffiltext{4}{Naval Research Laboratory, Code 7213, Washington, 
DC 20375-5320, USA}
\altaffiltext{5}{Center for Space Research, Massachusetts Institute of 
Technology, 77 Massachusetts Avenue, Cambridge, MA 02139, USA}
\altaffiltext{6}{Department of Astronomy, University of California, 
601 Campbell Hall, Berkeley, CA 94720-3411, USA}
\altaffiltext{7}{Max-Planck-Institut f\"ur extraterrestrische Physik, 
PO Box 1312, 85741 Garching bei M\"unchen, Germany}

\shorttitle{{X-Ray, UV and Optical Monitoring of SN~1979C}}
\shortauthors{Immler et~al.}

\begin{abstract}

We present results from observations of \sn\ with \X\ in X-rays and in the UV,
archival X-ray and \H\ data, and follow-up ground-based optical imaging.
The \X\ MOS spectrum shows best-fit two-temperature thermal
plasma emission characteristics of both the forward
($kT_{\rm high}=4.1^{+76}_{-2.3}$~keV) and reverse shock
($kT_{\rm low}=0.78^{+0.25}_{-0.17}$~keV) with no intrinsic absorption.
The long-term X-ray lightcurve, constructed from all X-ray data
available, reveals that \sn\ is still radiating at a flux level similar 
to that detected by \R\ in 1995, showing no sign of a decline over the 
last six years, some 16--23~yrs after its outburst. The high inferred 
X-ray luminosity ($L_{0.3-2}=8\times10^{38}~{\rm ergs~s}^{-1}$) is caused 
by the interaction of the SN shock with dense circumstellar matter,
likely deposited by a strong stellar wind from the progenitor with a high 
mass-loss rate of 
$\dot{M} \approx 1.5 \times 10^{-4}~M_{\odot}~{\rm yr}^{-1}~(v_{\rm w}/10~{\rm km~s}^{-1})$.
The X-ray data support a strongly decelerated shock, and show
a mass-loss rate history which is consistent with a constant progenitor 
mass-loss rate and wind velocity over the past $\gs16,000$~yrs 
in the stellar evolution of the progenitor.
We find a best-fit CSM density profile of $\rho_{\rm CSM} \propto r^{-s}$ 
with index $s \ls 1.7$ and high CSM densities 
($\gs 10^4~{\rm cm}^{-3}$) out to large radii from the site of the explosion
($r \gs 4 \times 10^{17}~{\rm cm}$).
Using \X\ Optical Monitor data we further detect a point-like optical/UV 
source consistent with the position of \sn, with $B, U$, and $UVW1$-band 
luminosities of $5, 7$, and $9 \times 10^{36}~{\rm ergs~s}^{-1}$, respectively.
The young stellar cluster in the vicinity of the SN, as imaged by the \H\ 
and follow-up ground-based optical imaging, can only provide a fraction 
of the total observed flux, so that a significant contribution to the output
likely arises from the strong interaction of \sn\ with dense CSM.

\end{abstract}


\section{Introduction}
\label{introduction}

The interaction of the outgoing supernova (SN) shock and ejecta with the 
circumstellar medium (CSM) can produce substantial amounts of X-ray emission.
The dominant X-ray production mechanism is the interaction of the SN
shock with the ambient CSM deposited either by a pre-SN stellar wind or 
non-conservative mass transfer to a companion. This interaction produces hot 
gas with a characteristic temperature in the range $T \sim 10^7$--$10^9$~K
(Chevalier \& Fransson 1994). Gas heated to such a high temperature produces 
radiation predominantly in the X-ray range. X-ray emission from this interaction 
is expected for all Type Ib/c and II SNe with substantial CSM established by 
the massive progenitors. The details of the CSM interaction are discussed in 
Fransson, Lundqvist \& Chevalier \cite{flc96} and references therein, while the
X-ray interaction luminosity is briefly summarized here (see also
Immler \& Lewin 2003): the thermal X-ray luminosity, $L_{\rm x}$, 
produced by the shock heated CSM, can be expressed as 
$L_{\rm x} = 4/(\pi m^2) \Lambda(T) \times (\dot{M}/v_{\rm w})^2\times(v_{\rm s} t)^{-1}$, where $m$ is the mean mass per particle ($2.1\times10^{-27}$~kg for a H+He 
plasma), $\Lambda(T)$ the cooling function of the heated plasma at temperature
$T$, $\dot{M}$ the mass-loss rate of the progenitor, $v_{\rm w}$ the speed
of the stellar wind blown off by the progenitor, and $v_{\rm s}$ the
speed of the outgoing shock. If X-ray spectroscopic data are available, 
$\Lambda(T)$ is known, and the interaction luminosity $L_{\rm x}$ at 
time $t$ after the outburst can be used to measure the ratio 
$\dot{M}/v_{\rm w}$. Assuming a constant SN shock velocity 
$v_{\rm s}$, the shock will reach a radius of $r=v_{\rm s}t$ 
from the site of the explosion at time $t$ after the explosion.

If the CSM density $\rho_{\rm csm}$ is dominated by a wind from the 
progenitor of the SN, the continuity equation requires 
$\dot{M} = 4\pi r^2 \rho_{\rm w}(r) \times v_{\rm w}(r)$ through a sphere 
of radius $r$, where $\rho_{\rm w}$ is the stellar wind density.
After the SN shock plows through the CSM, its density is 
$\rho_{\rm csm} = 4\rho_{\rm w}$ (Fransson, Lundqvist \& Chevalier 1996),
assuming no losses to cosmic ray acceleration.

Since each X-ray measurement at time $t$ is related to the corresponding 
distance $r$ from the site of the explosion, which has been reached 
by the wind at a time depending on $v_{\rm w}$, or the age of the wind 
$t_{\rm w} = t v_{\rm s}/v_{\rm w}$, we can use our measurements as a 
`time machine' to probe the progenitor's history over significant time scales.
Assuming that $v_{\rm w}$ and $v_{\rm s}$ did not change over $t_{\rm w}$, 
or in cases where the shock front deceleration is known from radio data,
we can even directly measure the mass loss rate back in time. 
Furthermore, integration of the mass-loss rate along the path of the 
expanding shell gives the mean density inside a sphere of radius $r$. 
For a constant wind velocity $v_{\rm w}$ and constant mass-loss rate $\dot{M}$, 
a $\rho_{\rm csm} = \rho_0 (r/r_0)^{-s}$ profile with $s=2$ is expected.

After the expanding shell has become optically thin, it is expected that 
emission from the SN ejecta itself, heated by the reverse shock, dominates 
the X-ray output of the interaction regions due to its higher emission
measure and higher density. Since the plasma temperature of the ejecta
is lower than that of the shocked CSM, a significant `softening'
of the X-ray spectrum is expected (e.g., Chevalier \& Fransson 1994;
Fransson, Lundqvist \& Chevalier 1996).

The number of detected X-ray SNe has substantially grown over the past
few years. With 22 detections to date\footnote{A complete list of
X-ray SNe and references are available at
http://lheawww.gsfc.nasa.gov/users/immler/supernovae\_list.html}
and high-quality X-ray spectra available from \X\ and \C\ observations, 
a wealth of new information has been gathered (see Immler \& Lewin 2003 
for a review article).

The signatures of circumstellar interaction in the radio, optical, and
X-ray regimes have been found for a number of core-collapse SNe, such 
as the Type II-L \sn\ (e.g., Weiler et~al.\ 1986, 1991; Fesen \& Matonick
1993; Immler et~al.\ 1998).
\sn\ was discovered in the optical near maximum light on April 19, 1979 
(Johnson 1979), and detected in X-rays 16~yrs after its outburst
with the \R\ HRI, with a luminosity of 
$L_{\rm 0.1-2.4}=1.3\times10^{39}~{\rm ergs~s}^{-1}$ (Immler et~al.\ 1998). 
The \R\ data imply a mass-loss rate of 
$\dot{M}\sim1\times10^{-4}~M_{\odot}~{\rm yr}^{-1}~(v_{\rm w}/10~{\rm km~s}^{-1})$,
similar to the mass-loss rates of other massive ($\gs 10M_{\odot}$) 
SN progenitors (e.g. SNe 1978K, 1986J, 1988Z, and 1998S)
and in agreement with the mass-loss rate inferred from \V\
radio observations 
($\dot{M}=1.2\times10^{-4}~M_{\odot}~{\rm yr}^{-1}~[v_{\rm w}/10~{\rm km~s}^{-1}]$;
Weiler et~al.\ 1991).
Earlier \E\ observations only gave an upper limit to the X-ray 
luminosity on days 64, 239, and 454 after the outburst, with $3\sigma$ 
upper limits of $1.8 \times 10^{40}~{\rm ergs~s}^{-1}$, 
$7.6 \times 10^{39}~{\rm ergs~s}^{-1}$ and 
$6.9 \times 10^{39}~{\rm ergs~s}^{-1}$, respectively
(Immler et~al.\ 1998). 
Since the \R\ HRI instrument does not provide energy information, and subsequent 
\A\ and \C\ observations (Kaaret 2001; Ray, Petre \& Schlegel 2001) lacked 
enough photon statistics for X-ray spectroscopy, there is no information 
about the X-ray spectral properties of \sn\ and thus the temperatures of the 
forward and reverse shock.

\sn\ was extensively monitored in the radio starting eight days after
maximum light, and showed an initial rate of decline of $t^{-0.7}$
(Weiler et~al.\ 1986, 2001), followed by a flattening or possibly brightening
at an age of $\approx 10$~yrs, while maintaining a relatively 
constant spectral index (Montes et~al.\ 2000; Weiler et~al.\ 2001). 
The results were interpreted as being due to a denser region in the CSM 
or a complex CSM structure which might have been modulated by the stellar 
wind of the progenitor's companion (Weiler et~al.\ 1992).

There are conflicting reports in the literature regarding the expansion
velocity of the ejecta. While Bartel \& Bietenholz \cite{bartel03} find a 
near-free expansion (i.~e., $E(T,V)=[{\rm const.}]$)
based on \VLBI\ data from $t=3.7$--$22~{\rm yrs}$
after the explosion, Marcaide et~al.\ \cite{marcaide02} report the 
detection of a strongly decelerated expansion. There is consensus,
however, that the initial expansion of the shock over the first
$6\pm2$~yrs was free, with $v_{\rm s} \approx 9,000~{\rm km~s}^{-1}$
(see Marcaide et~al.\ 2002 and references therein). The claimed onset 
of the deceleration to $v_{\rm s} \approx 6,200~{\rm km~s}^{-1}$
coincides with the flattening of the radio lightcurve and might be
indicative for the shock running into a dense CSM (Marcaide et~al.\ 2002).

Low-dispersion Keck optical spectra obtained at $t=14.8$~yrs and 17~yrs,
as well as \H\ UV spectra 17~yrs after the outburst revealed broad
$\approx$ 6,000~${\rm km~s}^{-1}$ emission lines, as well as blueward
peaks which are substantially stronger than those toward the red,
indicating internal dust extinction within the expanding ejecta.
Profile differences between  hydrogen and oxygen line emission
suggest two or more separate emitting regions, such as a H$\alpha$
bright shell and oxygen inner ejecta region (Fesen et~al.\ 1999).

\H\ imaging of the \sn\ environment showed the presence of a cluster 
of stars in the vicinity of \sn, containing both young
($\approx 4$--$8$~Myrs) blue stars, as well as older red supergiants
($\approx 20$~Myrs), and succeeded in identifying \sn\ as the brightest 
source in the optical and UV in that cluster
(van Dyk et~al.\ 1999). The data were further used to estimate the
mass of the progenitor to be $18\pm3~M_{\odot}$, in
agreement with a mass-estimate derived from radio data
($\gs 13~M_{\odot}$, Weiler et~al.\ 1991).

In this paper we report on \X\ X-ray and UV observations of \sn,
all archival X-ray data available, as well as archival \H\ images 
and new ground-based optical data. In \S \ref{data}
we describe the data and analysis thereof and report on the
X-ray spectrum, multi-mission X-ray lightcurve,  UV photometry,
and new follow-up ground-based optical imaging in \S \ref{results}. 
We discuss the results in the context of the CSM 
interaction model and compare them to other core-collapse SNe in 
\S \ref{discussion}, followed by a brief summary in \S \ref{summary}.

\section{Data Processing and Analysis}
\label{data}
SN~1979C and its host galaxy NGC~4321 (M100) were observed with \X\ as 
part of a GTO program on December 28, 2001, with the EPIC PN instrument
for 31.6~ks, and the EPIC MOS1 and MOS2 instruments for 36.2~ks
and 36.0~ks, respectively, using the medium filter.
Data processing and analysis were performed with SAS~6.1 and the latest
calibration constituents, as well as FTOOLS and own IDL routines.
Spectral fitting was performed using XSPEC 11.3.1.

In order to establish the long-term X-ray lightcurve of \sn\ we further
extracted and reduced all X-ray data available in the archives
(see \S \ref{other}). \X\ Optical Monitor data of SNe 1979C
(\S \ref{optical_monitor}), as well as archival \H\ and recent MDM 
observations (\S \ref{mdm}) were further used to estimate optical/UV fluxes.

\subsection{{\sl XMM-Newton} EPIC Data}
\label{epic}

Screening of the PN and MOS data for periods with a high 
background revealed contamination of the data at the end 
of the  observation period. Exclusion of these flaring periods gave
cleaned exposure times of 27.1~ks (PN) and 23.6~ks (MOS).

\begin{figure*}[p!]
\unitlength1.0cm
    \begin{picture}(5,9) 
\put(0,0){ \begin{picture}(8,8)
	\psfig{figure=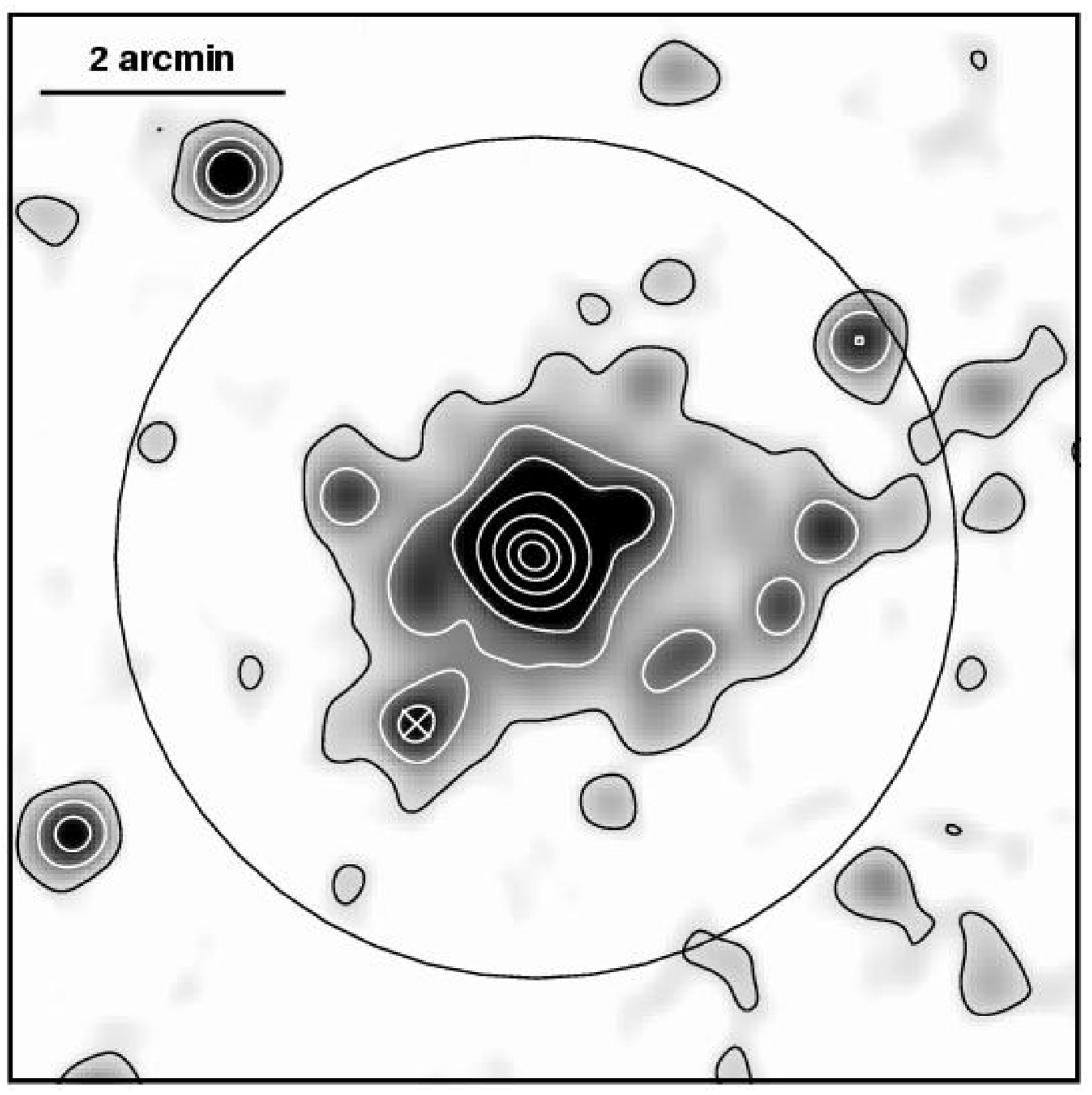,width=9cm,angle=0}
	\end{picture}
	}
\put(9.3,0){ \begin{picture}(8,8)
	\psfig{figure=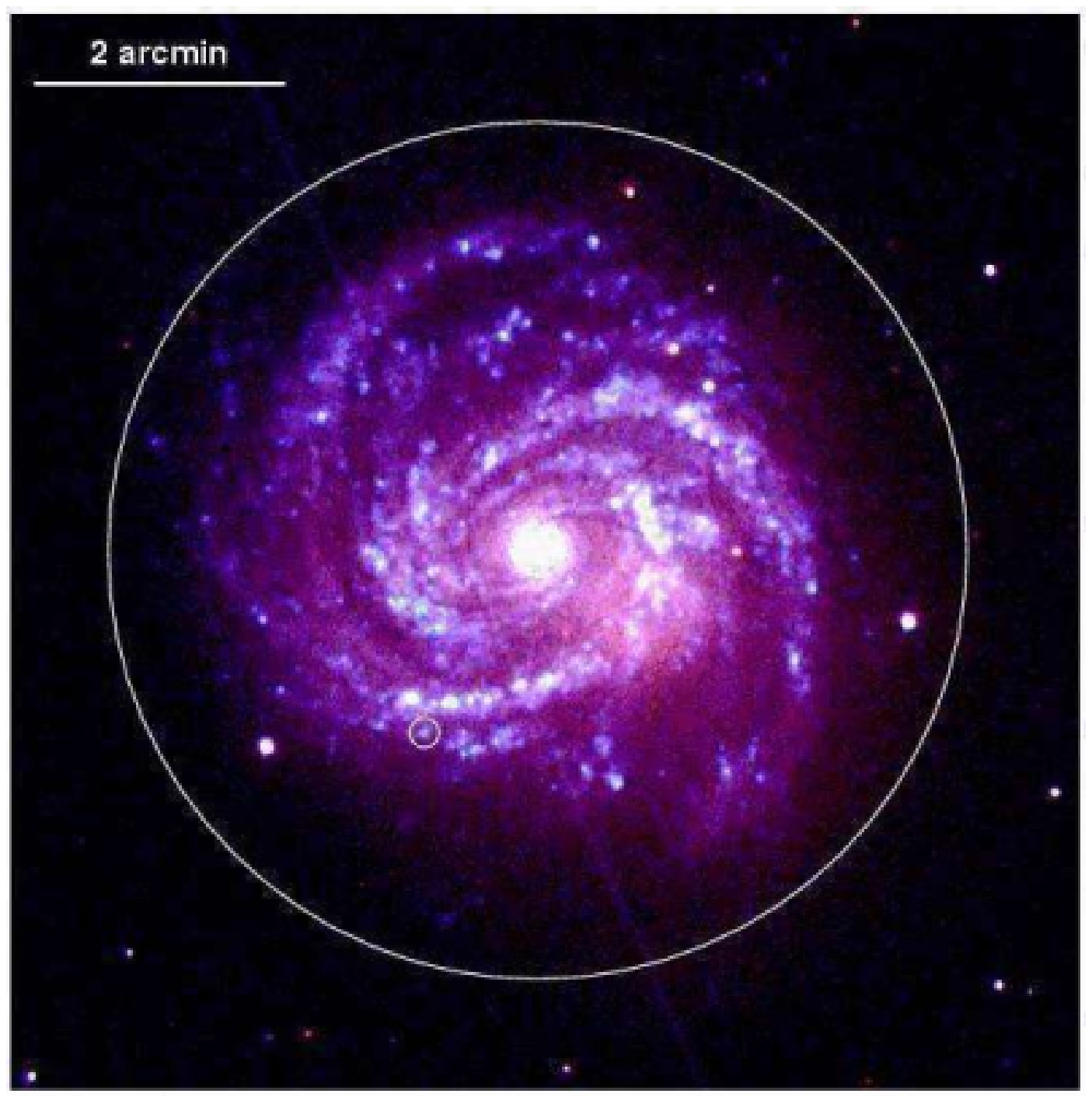,width=8.75cm,angle=0}
	\end{picture}
	}
    \end{picture}
\caption{{\bf Left-hand panel:} Merged \X\ EPIC MOS1+MOS2 image of M100 and 
\sn\ in the 0.3--10~keV band in logarithmic gray-scale. 
The image was adaptively smoothed to achieve a signal-to-noise ratio of $10$. 
Contour lines are $0.65$ (black), $1$, $1.5$, $3$, $6$, $10$, and 
$13\times10^{-5}~{\rm cts~s}^{-1}~{\rm pix}^{-1}$. The position of \sn\ 
is marked by a white cross. The black circle gives the $D_{25}$ ellipse of M100.
{\bf Right-hand panel:} $B$ (red), $U$ (green), and $UVW1$ (blue) Optical 
Monitor image of M100. The small white circle (radius $7\farcs5$) is centered on 
the radio position of \sn. The large white circle indicates the $D_{25}$ ellipse 
of M100 as in the left-hand panel.}
\label{image}
\end{figure*}

Inspection of the EPIC images further showed that the position of
\sn\ coincides with a chip gap on the PN detector, making a
reliable count rate estimate impossible. The PN data were therefore 
not used in the analysis and we will restrict ourselves to the
EPIC MOS data only. The MOS data were further screened using the
parameters ``FLAG=0'' and ``PATTERN $<$= 12'' and filtered to retain 
the 0.3--10~keV energy band only in order to avoid contamination from 
soft-proton flaring and uncertainties in the calibration at low energies.
The merged 0.3--10~keV band MOS1+MOS2 image of M100/SN~1979C is 
shown in the left-hand panel of Fig.~\ref{image}. 
MOS counts were extracted within a circle of radius $30''$ centered
on the radio position of \sn\ and corrected for the energy-enclosed 
aperture. The background was estimated locally from a source-free annulus 
with inner and outer radii of $45''$ and $55''$, respectively, to account
for residual diffuse emission from the host galaxy.

\subsection{{\sl XMM-Newton} Optical Monitor Data}
\label{optical_monitor}

Optical Monitor (OM) data were obtained in 16 individual exposures
in the $B$-band (effective wavelength 4,340~\AA), $U$-band (3,440~\AA), 
and $UVW1$-band (2,910~\AA), with on-source exposure times of 2.0~ks 
($B$), 1.8~ks ($U$), and 1.9~ks ($UVW1$). OM data were reprocessed using 
the SAS ``omichain'' task, and following the data analysis threads in the 
ABC Guide version 2.01\footnote{available from
http://heasarc.gsfc.nasa.gov/docs/xmm/abc/}. 

For imaging and source detection, the tracking history was created,
bad pixels were removed, a flat field generated, and images were 
corrected for ``modulo-8'' spatial fixed-pattern noise. Source detection 
was performed using the ``omdetect'' task, which employs a two-stage 
process to locate sources in OM Science Window images, i.e., determination 
of the background and consecutive island search (``box detect'') in which 
sets of pixels above the sigma significance cut-off are identified and grouped 
into individual objects.
Source counts were converted into instrument-bandpass magnitudes utilizing
the ``ommag'' command. Fluxes were converted from count rates using a 
step-by-step recipe provided by Alice Breeveld\footnote{available from
http://xmm.vilspa.esa.es/sas/documentation/watchout/uvflux.shtml}.
We also employed the interactive ``omsource'' task with various aperture
sizes to verify the magnitudes obtained with ``omichain'' and to check
for potential contamination of the OM photometry with nearby sources.

Comparison of the optical positions of six stars from the Guide Star 
Catalog with the centroid positions of the stars in the $B$, $U$, and 
$UVW1$ images revealed an offset of individual OM images of 
$\Delta{\rm R.A.}=0\farcs14$ and $\Delta{\rm Dec.}=-0\farcs95$ ($B$),
$\Delta{\rm R.A.}=0\farcs43$ and $\Delta{\rm Dec.}=-1\farcs75$ ($U$), and
$\Delta{\rm R.A.}=1\farcs01$ and $\Delta{\rm Dec.}=-2\farcs23$ ($UVW1$).
This offset was also visible in an overlay of the $B$, $U$, and $UVW1$ images.
After re-alignment of the images with shifts in R.A. and Dec. as quoted above, 
no systematic offset between the different images, and no statistically 
significant offset between the centroid positions of the stars with the 
Guide Star Catalog positions are observed. 

\begin{figure*}[h!]
\unitlength1.0cm
    \begin{picture}(5,11.5) 
\put(-2.8,0){ \begin{picture}(5,10)
	\psfig{figure=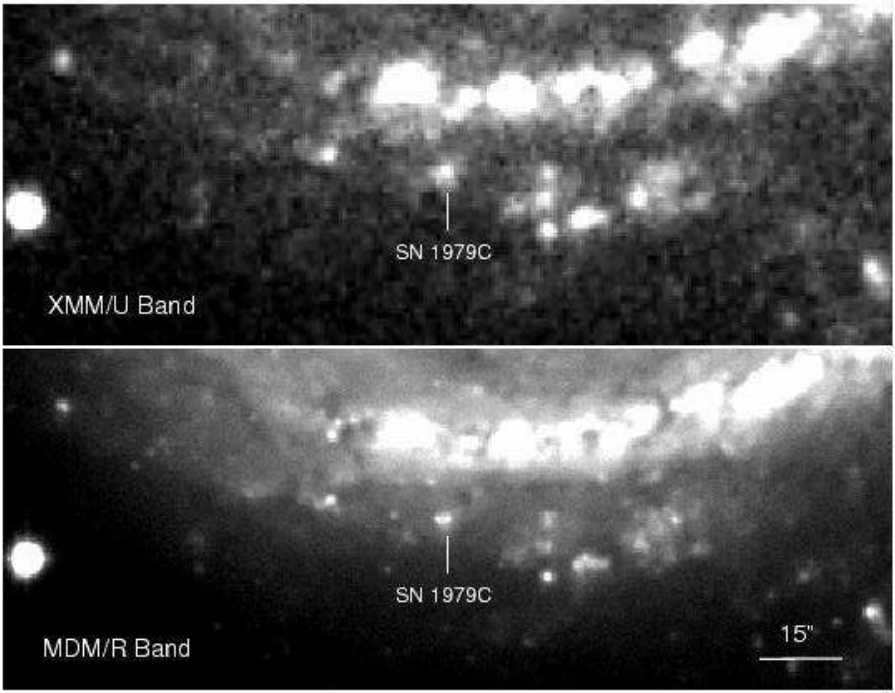,width=17.5cm,angle=-90}
	\end{picture}
	}
    \end{picture}
\vspace{-1.7cm}
\caption{
\X\ Optical Monitor $U$-band image (upper panel) and MDM 2.4-m Hiltner Telescope
$R$-band image of the southern spiral arm of M100 centered on the position of \sn.
The $15''$ scalebar in the lower panel applies to both panels.
} 
\label{spiral}
\end{figure*}

A three-color ($B$, $U$, $UVW1$) image of M100 and SN~1979C is shown in the 
right-hand panel of Fig.~\ref{image}. 
The circular enhancement in the south-western spiral arm of M100, 
most prominent in the $B$-band image (color-coded in red), is due to stray-light
from the bright nucleus of M100. Due to its large offset from the position
of \sn\ and local background extraction, however, it does not affect the 
\sn\ photometry. An \X\ OM $U$-band image of the region around the position 
of \sn\ is further shown in more detail in Fig.~\ref{spiral} (upper panel).

\subsection{Other X-Ray Data}
\label{other}
In addition to the \X\ data described above, we used archival data from 
\E\ (HRI), \R\ (HRI and PSPC), \A\ (SIS), and a recent \C\ (ACIS-S)
observation. All \E\ observations were merged into a single 41.2~ks
observation to increase the photon statistics.
Count rates from the archival \E\ HRI and \R\ HRI observations 
were taken from Immler et~al.\ \cite{immler98},
\C\ count rates were taken from Ray, Petre \& Schlegel \cite{ray01}
and Kaaret \cite{kaaret01}, and \A\ upper limits from Ray, Petre \& Schlegel 
\cite{ray01}. 
We further extracted source counts for each of these observations following 
standard analysis methods as described in the respective instrument 
handbooks using FTOOLS to check and reproduce published count rates.
Two previously unpublished \R\ HRI observations were analyzed as described 
in Immler et~al.\ \cite{immler98}.

A \R\ PSPC observation from December 8, 1998, was also used to establish the 
long-term X-ray lightcurve of \sn. This observation was selected
as part of the \R\ PSPC ``last-light'' campaign during which
the PSPC instrument was re-activated for a final set of observations
just days before \R\ was decommissioned. Inspection of the PSPC data showed 
that the position of \sn\ was sufficiently offset from the ``gain hole'' which 
emerged on the PSPC-B detector images taken during the short ``last-light'' campaign.
Exposure corrected counts were extracted from the position of \sn\
within a circle of radius $30''$ and corrected for the 100\% encircled
energy radius. The background was estimated locally within an annulus
of inner and outer radii of $1'$ and $1\farcm3$. The net exposure time
of the observation is 7.8~ks. A log of all X-ray observations utilized 
in this paper is given in Table~\ref{tab1}.

\subsection{Follow-up Ground-based Optical Imaging}
\label{mdm}

H$\alpha$ and $R$(Harris)-band images were obtained at the MDM Observatory
on December 19 and 20, 2004 (epoch = 2004.97 $\simeq$ 2005.0), using the 
Hiltner 2.4~m telescope and the Columbia 8K CCD camera (Crotts 2001).
The images were used to investigate further the nature of the source detected 
at the \sn\ site, and to compare more directly its current optical flux level 
with that seen in previous late-time optical detections (Fesen et al. 1999; 
van Dyk et al.\ 1999), 
Two 1,000~s $R$-band and three 1,000~s H$\alpha$ exposures binned $2 \times 2$ 
yielding an image scale of 0.404 arcsec/pixel were taken under poor to fair 
conditions. The H$\alpha$ image had poor image quality ($3''$ FWHM) while 
the $R$-band images were fair to good ($1\farcs2$--$1\farcs4$ FWHM). 
These data were reduced, co-added and cosmic ray corrected using standard 
IRAF software routines. The MDM $R$-band image of the region around the 
position of \sn\ is shown in Fig.~\ref{spiral} (lower panel).

As a means for comparison of current optical flux levels, we also retrieved and
reprocessed the archival {\sl HST} WFPC-PC1 images of the \sn\ region
(c.f. van Dyk et al.\ 1999). These data consist of several filter images taken
July 29, 1996, using the WFPC2 camera with the \sn\ site centered in the high
resolution Planetary Camera (PC1; 0.046 arcsec/pixel). 
Two 500~s broad-band red (F675W), three 1,300~s H$\alpha$ (F658N), 
two 1,200~s $B$-band (F439W), and two 1,300~s $U$-band (F336W) filter images 
were co-added and cosmic rays removed via high pixel value rejection procedure.
A few obvious residual cosmic ray contaminated pixels on the summed F675W image
were also removed manually using the IRAF software package ``tv.imedit''.
All optical/UV observations utilized in this paper are listed in Table~\ref{tab2}.

\section{Results}
\label{results}

\subsection{X-Ray Data}
\label{results_xmm}

\begin{figure*}[h!]
\unitlength1.0cm
    \begin{picture}(5,8) 
	\psfig{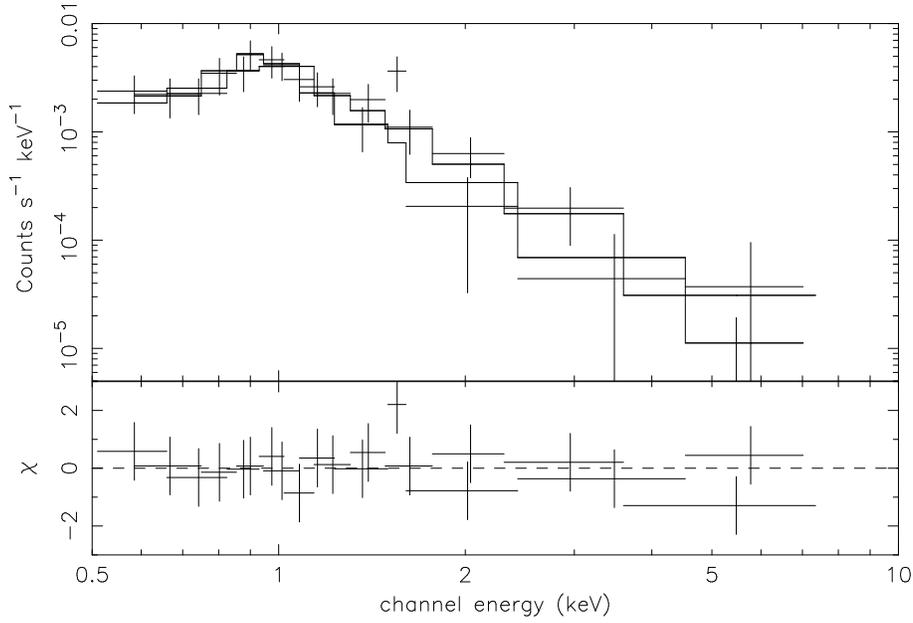}
     \end{picture}
\caption{EPIC MOS1 and MOS2 X-ray spectra of \sn\ (upper panel). 
The lower panel shows the residuals of the fits in units of $\sigma$
for the best-fit spectral model consisting of a two-component thermal 
plasma emission model (see \S \ref{results_xmm}).
\label{spectrum}}
\end{figure*}

\sn\ is detected in the MOS images with a count rate of
($3.6\pm0.5) \times 10^{-3}~{\rm cts~s}^{-1}$, 23~yrs after its outburst.
Simultaneous spectral fitting of the MOS data in the 0.5--8~keV band
gives a best-fit (reduced $\chi^2_{\rm r} = 0.65$ for 18 degrees of freedom) 
two-temperature plasma emission model (XSPEC model `MEKAL') with
$kT_{\rm low}=0.77^{+0.17}_{-0.19}$~keV and
$kT_{\rm high}=2.31^{+1.95}_{-0.66}$~keV, and a best-fit absorbing column
density of $N_{\rm H} = 2.53^{+4.96}_{-2.53} \times 10^{20}~{\rm cm}^{-2}$
consistent with the Galactic foreground column density
($N_{\rm H} = 2.39 \times 10^{20}~{\rm cm}^{-2}$; Dickey \& Lockman 1990). 
Element abundances cannot be constrained due to the low photon statistics. 
In order to avoid contamination of the spectra with unresolved and hard
XRBs within the disk of M100, we further fitted the spectrum in the
0.5--2~keV band, where the bulk of emission from the reverse shock is
expected. A best-fit two-temperature thermal plasma emission model 
($\chi^2_{\rm r} = 0.61$, ${\rm d.o.f}=8$) gives
$kT_{\rm low}=0.78^{+0.25}_{-0.17}$~keV and
$kT_{\rm high}=4.1^{+76}_{-2.3}$~keV, and an absorbing column
density of $N_{\rm H} = 0$--$1.01 \times 10^{21}~{\rm cm}^{-2}$.
Adopting this spectral template gives 0.3--2~keV and 2--8~keV
fluxes of $f_{0.3-2}=2.31 \times 10^{-14}~{\rm ergs~cm}^{-2}~{\rm s}^{-1}$ and 
$f_{2-8}=1.81 \times 10^{-14}~{\rm ergs~cm}^{-2}~{\rm s}^{-1}$, respectively.
The soft component contributes 64\% to the total emission, while the
hard component accounts for the remaining 36\% of the 0.3--2~keV band flux.
The MOS spectra and best-fit model are shown in Fig.~\ref{spectrum}. 

A single component thermal plasma spectrum gives best-fit 
$kT=0.60^{+0.12}_{-0.19}$~keV and an absorbing column density of
$N_{\rm H} = 6.5^{+2.0}_{-1.9} \times 10^{21}~{\rm cm}^{-2}$
($\chi^2_{\rm r} = 1.29$, ${\rm d.o.f}=11$). 
Using these spectral properties gives a soft-band flux of 
$f_{0.3-2}=1.69 \times 10^{-14}~{\rm ergs~cm}^{-2}~{\rm s}^{-1}$.
A thermal bremsstrahlung spectrum with $kT=0.53^{+1.40}_{-0.33}$~keV 
($\chi^2_{\rm r} = 0.91$, ${\rm d.o.f}=11$) cannot be statistically ruled out, 
but shows significantly larger systematical offsets from the observed 
spectrum, especially at lower energies. 
Adopting this spectral template gives a soft-band flux of 
$f_{0.3-2}=2.11 \times 10^{-14}~{\rm ergs~cm}^{-2}~{\rm s}^{-1}$.
Other spectral models, such as a power-law ($\Gamma=4.48^{+0.53}_{-0.48}$;
$\chi^2_{\rm r} = 0.87$, ${\rm d.o.f}=17$) or a non-equilibrium ionization model 
($kT=73.4^{+6.5}_{-33.0}$~keV; $\chi^2_{\rm r} = 0.93$, ${\rm d.o.f}=16$)
cannot be statistically ruled, but give unreasonable physical parameters.

Using the above two-temperature thermal plasma spectral template and assuming 
a distance of $17.1$~Mpc (Freedman et~al.\ 1994), we infer soft- and 
hard-band X-ray luminosities of
$L_{0.3-2}=8 \times 10^{38}~{\rm ergs~s}^{-1}$ and 
$L_{2-10}=3 \times 10^{38}~{\rm ergs~s}^{-1}$, respectively.

\begin{figure*}[h!]
\unitlength1.0cm
	\begin{picture}(5,8) 
	\psfig{figure=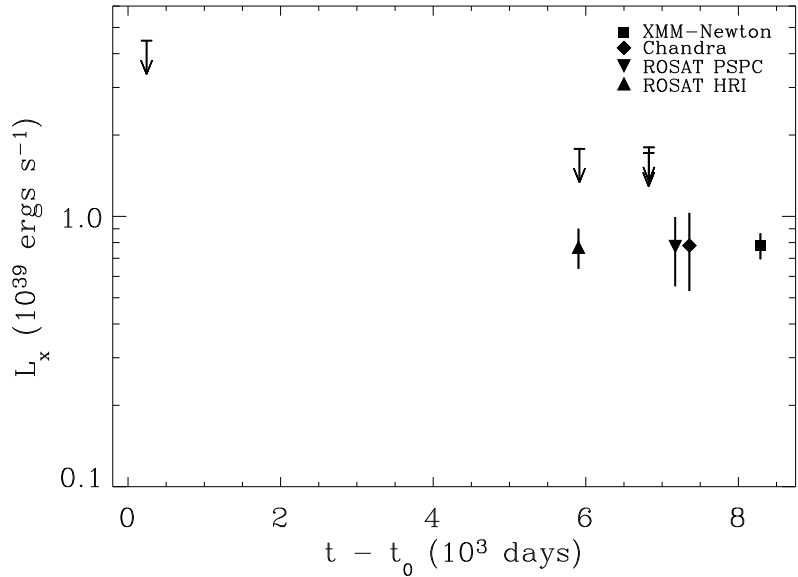,width=12cm,angle=0}
     \end{picture}
\caption{Soft (0.3--2~keV) band X-ray lightcurve of
\sn\ based on \E, \R, \A, \C, and \X\ data. Upper limits are from \E\ HRI,
\R\ HRI and \A\ SIS observations. Time is given in units of $10^3$ days after 
the outburst of \sn\ (see Table~1). \label{lightcurve}}
\end{figure*}

We also analyzed the multi-mission data set to construct the long-term
X-ray lightcurve of \sn, using the same spectral template to convert
count rates into fluxes and luminosities.
The long-term X-ray lightcurve is shown in Fig.~\ref{lightcurve}.
We calculated the mass-loss rate of the progenitor as a function 
of age of the stellar wind assuming a constant shock velocity of 
9,200~${\rm km~s}^{-1}$ (Marcaide et~al.\ 2002). 
An effective (0.3--2~keV band) cooling function of
$\Lambda = 3 \times 10^{-23}~{\rm ergs~cm}^3~{\rm s}^{-1}$ for an optically 
thin thermal plasma with a temperature of $10^7$~K was adopted, which 
corresponds to the temperature inferred from the MOS spectra. 
Adopting different plasma temperatures in the range 0.5--1~keV
would lead to changes in the emission measure of $\approx 10\%$.
This error in the emission measure is included in calculations of
the CSM number densities (\S~\ref{xray-data}).
Key physical properties of \sn\ are listed in Table~\ref{tab3}.

While the mass-loss rate history (see Table~\ref{tab3} and left-hand panel 
of Fig.~\ref{massloss})
indicates a slight increase in the mass-loss rate toward a larger stellar
wind age, the results are not statistically significant.
We further calculated the mass-loss rate history assuming a 
strongly decelerated expansion of the shock of 
9,200~${\rm km~s}^{-1}$ for a stellar wind age of $t_{\rm w}<6~{\rm yrs}$, 
6,200~${\rm km~s}^{-1}$ for $6~{\rm yrs}<t_{\rm w}<14~{\rm yrs}$, and
5,400~${\rm km~s}^{-1}$ for $t_{\rm w}>20~{\rm yrs}$,
corresponding to a deceleration $R \propto t^m$ with a deceleration
parameter of $m=0.6$ (Marcaide et~al.\ 2002) to approximate a continually 
decelerating shock. The results are shown in the right-hand panel of 
Fig.~\ref{massloss}. The introduction of a deceleration leads to a 
constant mass-loss rate of
$\dot{M}\approx 1.5 \times 10^{-4}~M_{\odot}~{\rm yr}^{-1}~(v_{\rm w}/10~{\rm km~s}^{-1})$.

We computed the CSM number density in the vicinity of the SN
by integrating the mass-loss rate within the sphere probed by the
CSM interaction. The post-shock CSM number densities, $n = \rho_{\rm csm}/m$, 
for the different radii $r=v_{\rm s}t$ corresponding to the dates of the 
observations, are listed in Table~\ref{tab3}.
An overall decrease in the CSM density is observed from
$n=(1.17\pm0.32) \times 10^4~{\rm cm}^{-3}$ at log\,$r[{\rm cm}]=17.50$ (\R\ HRI) to
$n=(0.85\pm0.19) \times 10^4~{\rm cm}^{-3}$ at log\,$r[{\rm cm}]=17.59$ (\X), albeit not
statistically significant. While the \R\ PSPC and \C\ data at log\,$r[{\rm cm}]=17.52$
($n=[1.66\pm0.63] \times 10^4~{\rm cm}^{-3}$) and $17.53$~cm
($n=[1.51\pm0.56] \times 10^4~{\rm cm}^{-3}$) are slightly higher, they are 
consistent with the earlier \R\ HRI and later \X\ data within the errors. 
The CSM density profile will be discussed in more detail in \S\ \ref{xray-data}.

\subsection{{\sl XMM-Newton} Optical Monitor Data}

An optical/UV source is found at the position of \sn\
in the $B$, $U$, and $UVW1$ bands, with magnitudes of
$m_{\rm B}=(19.1 \pm 0.1)$~mag,
$m_{\rm U}=(18.0 \pm 0.1)$~mag, and
$m_{\rm UVW1}=(17.6 \pm 0.1)$~mag, and source detection significances
of $17.1\sigma$, $28.1\sigma$, and $34.5\sigma$, respectively,
within an aperture of $12$~pix ($4\farcs1$~FWHM). The source is
flagged as point-like by the source detection algorithm.
Magnitudes were converted into fluxes with Vega as a reference
(Vega has magnitudes of $m_{\rm B}=0.030$~mag, and
$m_{\rm U}=m_{\rm UVW1}=0.035$~mag), and a relation of 
$m_{\rm Vega} - m_{\rm SN1979C} = -2.5~{\rm log}~(f_{\rm Vega}/f_{\rm SN1979C})$,
and Vega fluxes in the OM $B$, $U$, and $UVW1$ filters of
5.97, 3.15, and $3.73 \times 10^{-9}~{\rm ergs~cm}^{-2}~{\rm s}^{-1}~{\rm \AA}^{-1}$.
We obtain $B$, $U$, and $UVW1$ band fluxes for \sn\ of 1.4, 2.0, 
and $3.3 \times 10^{-16}~{\rm ergs~cm}^{-2}~{\rm s}^{-1}~{\rm \AA}^{-1}$, respectively.

In order to minimize contamination of the optical/UV with potential
unresolved neighboring sources, we also extracted OM counts within 
smaller apertures of 3 pixel radius ($2\farcs9$) from the centroid position
of \sn. Using this small aperture we obtain $m_{\rm UVW1}=(17.9 \pm 0.1)$~mag,
$f_{\rm UVW1}=2.6 \times 10^{-16}~{\rm ergs~cm}^{-2}~{\rm s}^{-1}~$\AA$^{-1}$, and
$L_{\rm UVW1}=9.0 \times 10^{36}~{\rm ergs~s}^{-1}$.

\subsection{$R$-Band and H$\alpha$ Imaging}

Our 2005.0 MDM $R$-band image of the \sn\ region in M100 is shown in 
Figs.~\ref{spiral} and \ref{optical}, along with the 2002.0 \X\ OM $U$-band 
image, and the 1996.6 {\sl HST} WFPC-PC1 red (F675W), H$\alpha$ (F658N), 
$B$-band (F439W), and $U$-band (F336W) images.
The MDM $R$-band image shows a bright source at the position of \sn\ with a strength
that suggests little change in $R$-band luminosity from that seen in the 1996.6
{\sl HST} F675W images.  Although the WFPC2 F675W and $R$(Harris) bandpasses are
different in wavelength coverage (F675W bandpass: 6,000--7,500~\AA; $R$(Harris)
bandpass: 5,600--8,200~\AA), they are both about equally sensitive to the
strong and broad emission lines of [O~I] $\lambda\lambda$6300,6364, H$\alpha$,
and [O~II] $\lambda\lambda$7319,7330 which dominate the late-time spectrum of
\sn\ (Fesen et al.\ 1999). This permits one to roughly equate these two
filter images for purposes of comparing total emission line levels in \sn.

Relative flux measurements on the WFPC2-PC1 F675W and MDM $R$-band images of the
\sn\ source and several surrounding point sources, including the bright source
$2\farcs0$ nearly due east, were taken and indicate SN~1979C's 2005.0 
6,000--7,500~\AA\ flux has remained constant to $\pm$0.25~mag from its 1996.6 level.
Concerning the current level of H$\alpha$ flux, the poor image quality of the
MDM H$\alpha$ image, differences in filter passbands between the WFPC2/F658N
($\lambda_{0}$ = 6,591~\AA; FWHM = 29~\AA) and the MDM 8K/H$\alpha$
filter ($\lambda_{0}$ = 6,565~\AA; FWHM = 80~\AA), plus uncertainty as to the
association of a strong, narrow H$\alpha$ at the rest frame velocity of M100 ($v$
= 1,570 km s$^{-1}$) with SN/CSM interaction make a definitive statement difficult.
Nonetheless, the ground-based image data do suggest that considerable H$\alpha$
emission currently exists at the \sn\ site at a level consistent with that 
seen in the 1996.0 WFPC2-PC1 image (see Fig.~\ref{optical}, right-hand panel).

\section{Discussion}
\label{discussion}

\subsection{X-Ray Data}
\label{xray-data}

\sn\ is detected at a high flux level 23~yrs after its outburst, with no
indication of a decline over the last six years.
This is in stark contrast to most other X-ray 
SNe which typically show X-ray and radio rates of decline of 
$L_{\rm x} \propto t^{n}$ with index $n$ in the range $0.3$--$1.4$ 
(see Immler \& Lewin 2003 and Sramek \& Weiler 2003 for review articles 
of the X-ray and radio rates of decline). The only SN which shows 
a similar persistent X-ray lightcurve is SN~1978K (Schlegel et~al.\
1999, 2004). Clearly, the high inferred flux is indicative of large
amounts of shocked CSM and a high mass-loss rate in the scenario
of the CSM being deposited by the progenitor. 

\begin{figure*}[h!]
\unitlength1.0cm
    \begin{picture}(5,7) 
\put(-1,0){ \begin{picture}(5,5.9)
	\psfig{figure=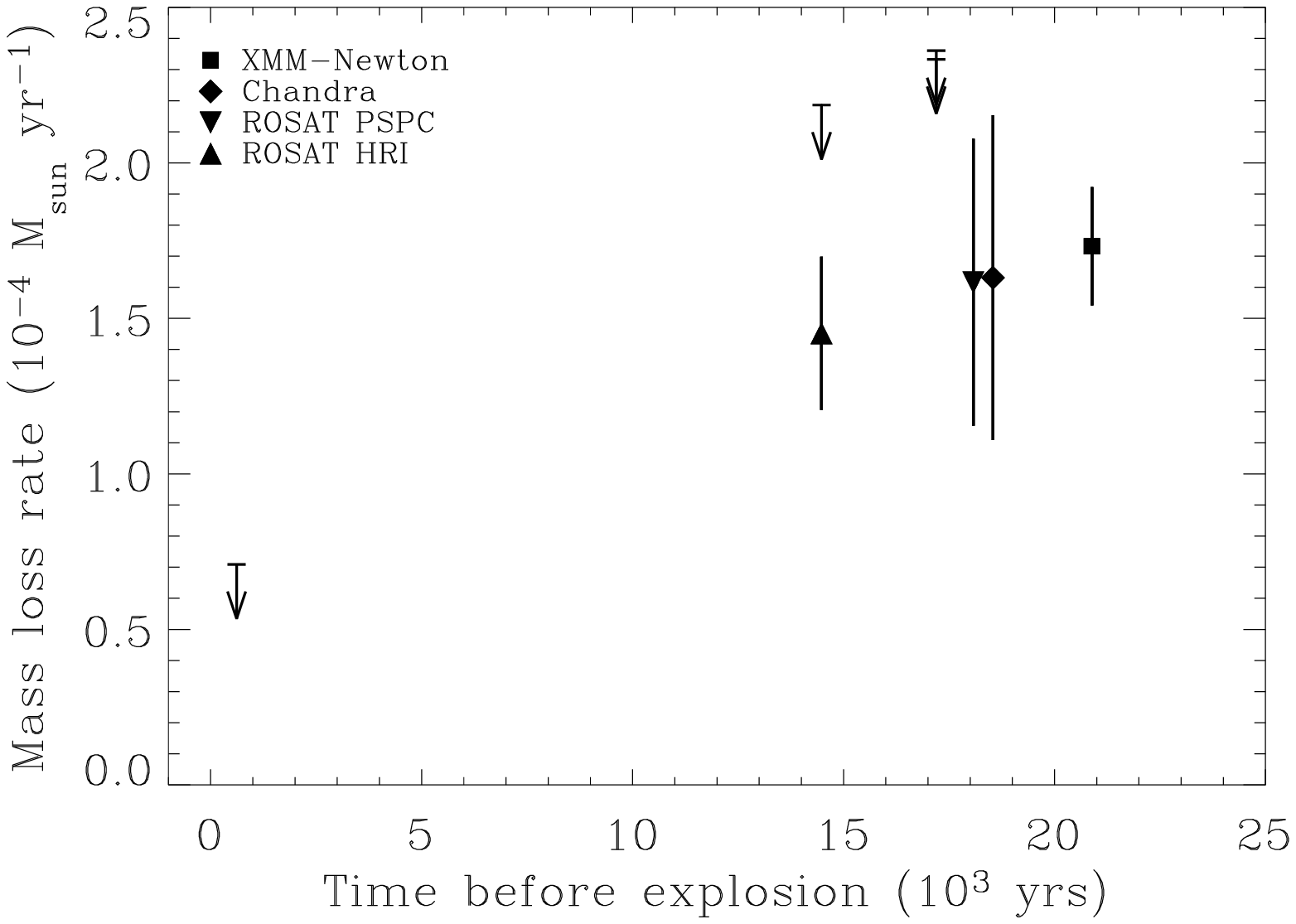,width=10cm,angle=0}
	\end{picture}
	}
\put(8.5,0){ \begin{picture}(5,5.9)
	\psfig{figure=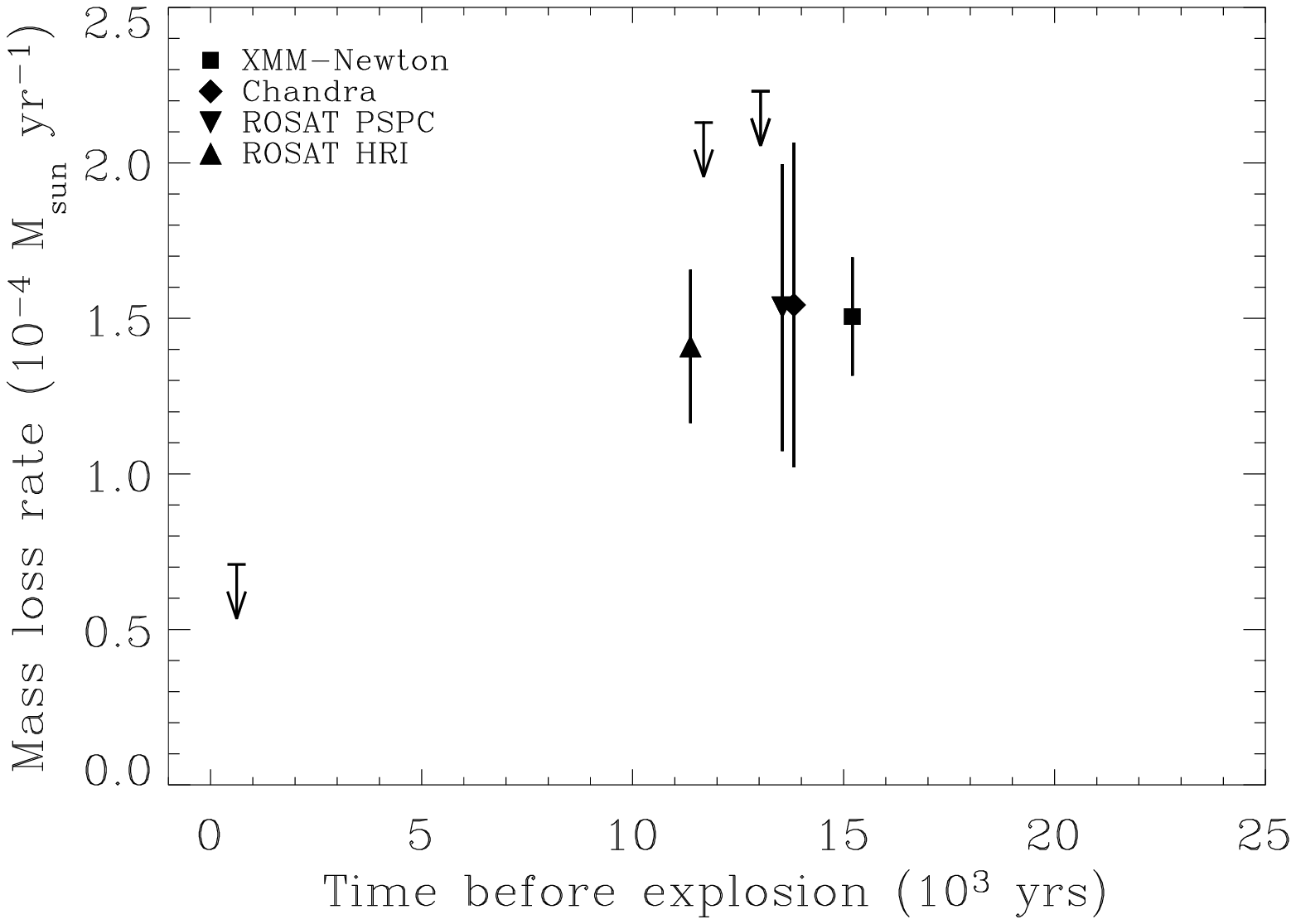,width=10cm,angle=0}
	\end{picture}
	}
    \end{picture}
\caption{{\bf Left-hand panel:}  Mass-loss rate history of \sn\ as a function of 
the stellar wind age in units of $10^3$ years before the explosion for a
constant shock velocity (see Table~1 and \S \ref{results}).
{\bf Right-hand panel:} Mass-loss rate history of \sn\ as a function of 
the stellar wind age in units of $10^3$ years before the explosion for
a decelerated shock (see \S \ref{results_xmm}).} 
\label{massloss}
\end{figure*}

Furthermore, our analysis may indicate a possible increase of the 
mass-loss rate for larger stellar wind ages (see left-hand panel of 
Fig.~\ref{massloss}) if we assume a constant velocity forward shock.
In order to check whether incorrect assumptions in the spectral properties
of \sn\ might mimic such an increase of the mass-loss rate, we investigated
the effect of the assumed spectral parameters used for the conversion
of source counts into fluxes. However, given our observed MOS count rates,
we cannot find spectral templates with realistic plasma temperatures
which could account for the apparent increase in the mass-loss rate.
In fact, if we adopt a spectral template which only includes a hard thermal 
component with $10$~keV for the forward shock emission for the earlier data 
for which no spectral information is available, the inferred flux would 
be $37\%$ {\it lower} when compared to our 
two-temperature thermal emission model, which takes the emergence of the 
(softer) reverse shock into account. We therefore can not reproduce a
flat mass-loss rate history by changing the plasma temperatures of
the shocked CSM alone. 

Instead, we find that the mass-loss rate history 
is best represented by a constant stellar wind velocity and constant 
mass-loss rate by assuming that the shock front experienced a strong 
deceleration. Adopting deceleration parameters derived from \VLBI\ imaging 
(with 9,200~${\rm km~s}^{-1}$ for $t_{\rm w}<6~{\rm yrs}$, 
6,200~${\rm km~s}^{-1}$ for $6~{\rm yrs}<t_{\rm w}<14~{\rm yrs}$, and
5,400~${\rm km~s}^{-1}$ for $t_{\rm w}>20~{\rm yrs}$; Marcaide et~al.\ 2002)
gives a flat (${\rm slope}=0$) mass-loss rate history (see right-hand panel 
of Fig.~\ref{massloss}). This indicates that the assumed deceleration
parameters are indeed plausible, and contradict claims for a free expansion 
based on the same \VLBI\ data (Bartel \& Bietenholz 2003). 

The early upper limit from the merged \E\ data 0.66~yrs after the outburst, 
however, is in conflict with
a constant stellar wind velocity and constant mass-loss rate model
(see Fig.~\ref{massloss}). Similarly to \sn, an early \R\ upper limit
to the X-ray flux of the SN Ic~1994I on day 52 has been reported to be 
in conflict with the X-ray evolution in terms of a simple power-law, 
and with a constant mass-loss rate and wind velocity (Immler et~al.\ 2002).
This is indicative that a simple power-law model for the X-ray rate of
decline might be incomplete for describing the early epoch in the CSM
interaction. Assuming an initial exponential rise of the X-ray luminosity 
after the outburst (at time $t_0$) and a subsequent power-law decline with
index $s$, we can parameterize the X-ray evolution as 
$f_{\rm x} \propto (t-t_0)^{-s} \times e^{-\tau}$ with $\tau \propto (t-t_0)^{-\beta}$. 
This model has been successfully used to describe the time dependence of the 
radio emission of SNe (Weiler et al.\ 1986) and the X-ray emission of SN~1994I
(Immler et~al.\ 2002). In this representation the external absorption of the 
emission is represented by the $e^{-\tau}$ term (`optical depth') and the 
time-dependence of the optical depth is parameterized by the exponent $\beta$. 
The rise of the X-ray flux could be either due to a decreasing absorption by 
intervening material along the line-of-sight to the X-ray emitting hot gas or 
could simply indicate a non-production of soft X-rays at early times.
Unfortunately, the lack of high-quality data for \sn\ at early times precludes 
a more detailed investigation. The low inferred mass-loss rate around 0.66~yrs
after the outburst of \sn, however, could be due to such an exponential increase
in the X-ray emission. The data can therefore be reconciled with a constant 
mass-loss rate history if we assume an initial rise of the emission, which
is observationally not further constrained.

Long-term X-ray lightcurves and high-quality spectra are also
available for SNe 1993J (Immler et~al.\ 2001; Zimmermann \& Aschenbach
2003) and 1978K (Schlegel et~al.\ 2004). While the overall X-ray lightcurve
of SN~1993J is best described by a $t^{-0.3}$ rate of decline,
significant `bumps' around day $\approx$ 1,000 are observed, which might be
due to a change in the slope of the CSM density profile
(Zimmermann \& Aschenbach 2003). \VLBI\ imaging revealed that the expanding 
shock region has a circular morphology, which is indicative of a smooth CSM.
Based on the X-ray and radio data, a significant change in the
mass-loss rate of the SN~1993J progenitor was discovered, with a
continuous decrease from $\dot{M}=4\times10^{-4}~M_{\odot}~{\rm yr}^{-1}$
to $4\times10^{-5}~M_{\odot}~{\rm yr}^{-1}$ during the late stages of 
the evolution (Immler et~al.\ 2001; van Dyk et~al.\ 1994). 
This evolution has been interpreted as a transition
in the progenitor's evolution from the red to the blue supergiant phase 
during the last $\approx 10,000$~yrs of the evolution (Immler et~al.\ 2001).
While a deceleration of the shock could not account for the change in the
mass-loss rate history for SN~1993J, correction for a strongly
decelerated shock produces a flat mass-loss rate history for \sn\ with 
$\dot{M} \approx 1.5 \times 10^{-4}~M_{\odot}~{\rm yr}^{-1}~(v_{\rm w}/10~{\rm km~s}^{-1})$.
The bumpy CSM structure inferred for \sn\ based on our analysis might
be largely explained by systematic and statistical uncertainties in the
X-ray flux, especially when the cross-calibration uncertainties between
\E, \R, \C, and \X\ are taken into account (which would add an
additional $\ls10$\% to the errors quoted in this paper;
see Snowden 2002 and the EPIC Calibration Documentation\footnote{available at 
http://xmm.vilspa.esa.es/external/xmm\_sw\_cal/calib/index.shtml}).
We therefore conclude that \sn\ does not show any evidence for a
change in the stellar wind parameters.

The X-ray spectrum of \sn\ is best described by a two-temperature
thermal plasma emission model with $kT_{\rm low}=0.78^{+0.25}_{-0.17}$~keV, 
$kT_{\rm high}=4.1^{+76}_{-2.3}$~keV, accounting for 64\% and 36\% of 
the total (0.3--2~keV) flux, respectively, and no intrinsic absorption.
Single-component fits cannot be statistically excluded and give a similar
temperature ($kT=0.60^{+0.12}_{-0.19}$~keV for a thermal plasma fit;
$kT=0.53^{+1.40}_{-0.33}$~keV for a thermal bremsstrahlung fit) and flux
compared to the soft component of the two-temperature fit. The hard and soft 
spectral components are likely due to emission from the forward (hard component)
and reverse shock (soft component). While minor contamination of the spectrum 
with unresolved XRBs within the disk of M100 cannot be conclusively excluded, 
spectral analysis of various point-source free regions within the $D_{25}$ 
ellipse of M100 give plasma temperatures significantly different from 
the spectral components above ($kT=[0.13\pm0.08]$~keV) and likely originate
in hot plasma within or above the disk of M100.

Similarly to \sn, a best-fit two-temperature thermal plasma model
with $kT_{\rm low}=(0.34\pm0.04)$~keV and $kT_{\rm high}=(6.54\pm4)$~keV
was observed for SN~1993J based on high-quality \X\ spectroscopy
eight years after its outburst (Zimmermann \& Aschenbach 2003). 
While these temperatures differ from the plasma temperatures found for \sn\ 
($kT_{\rm low}=0.78^{+0.25}_{-0.17}$~keV, $kT_{\rm high}=4.1^{+76}_{-2.3}$~keV)
the overall temperature evolution for SN~1993J with a softening
over time is consistent with the spectrum of \sn\ at this late stage.
Since no time-resolved spectroscopy is available for \sn, we used
the best-fit spectral template of the \X\ MOS data to convert count
rates into fluxes, luminosities, and compute mass-loss rates and
CSM densities. Changes in the spectral properties, however, do not
have a significant impact on the main results from this analysis
(see \S\ \ref{epic}).

\C\ data of SN~1998S, which was observed to have a similarly high 
mass-loss rate ($2\times10^{-4}~M_{\odot}~{\rm yr}^{-1}$)
also showed a softening of its spectrum from $10.4$~keV at day 678
after its outburst to $8.0$~keV on day 1,048 for a thermal plasma
fit with Galactic absorption only (Pooley et~al.\ 2002).
A more recent \X\ observation revealed the emergence of a soft
($\sim0.8$~keV) component, probably from the reverse shock, in addition 
to the harder component observed with \C\ (Immler \& Lewin 2003).

In terms of spectral properties, \sn\ shows a striking resemblance
to SN~1978K. At an age of $24.2$~yrs, the X-ray spectrum of SN~1978K
as observed with \X\ was best-fitted with a two-temperature thermal
plasma spectrum ($kT_{\rm low}=0.61^{+0.04}_{-0.05}$~keV,
$kT_{\rm high}=3.16^{+0.44}_{-0.40}$~keV;
Schlegel et~al.\ 2004). Similarly to \sn, SN~1978K shows no decline
over the observed period and has been detected at a persistently high X-ray 
luminosity ($L_{0.5-2}=1.5\times10^{39}~{\rm ergs~s}^{-1}$), which places
both SNe into the category of strongly interacting with dense ambient CSM.

\begin{figure*}[h!]
\unitlength1.0cm
	\begin{picture}(5,10) 
	\psfig{figure=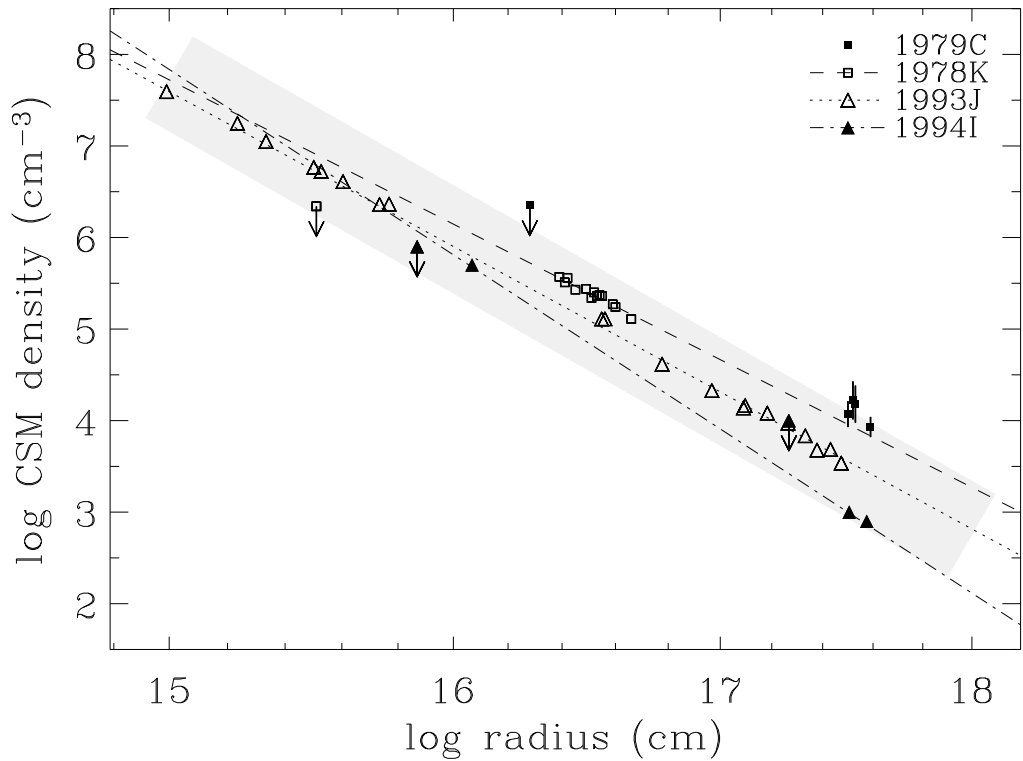,width=16cm,angle=0}
     \end{picture}
\caption{Circumstellar matter density profiles as a function of SN shell 
expansion radii for SNe 1994I (filled triangles; Immler et~al.\ 2002), 
1993J (open triangles; Immler et~al.\ 2001), 1978K (open box; this work), 
and 1979C (filled box; this work). Error bars for \sn\ are $\pm1\sigma$.
Best-fit CSM density profiles 
of $\rho_{\rm csm} \propto r^{-s}$ with indices $s=-1.9$ (SN~1994I), 
$s=-1.6$ (SN~1993J), and $s=-1.0$ (SN~1978K) are drawn as lines.
\label{rho}}
\end{figure*}

\begin{figure*}[h!]
\unitlength1.0cm
	\begin{picture}(5,11) 
	\psfig{figure=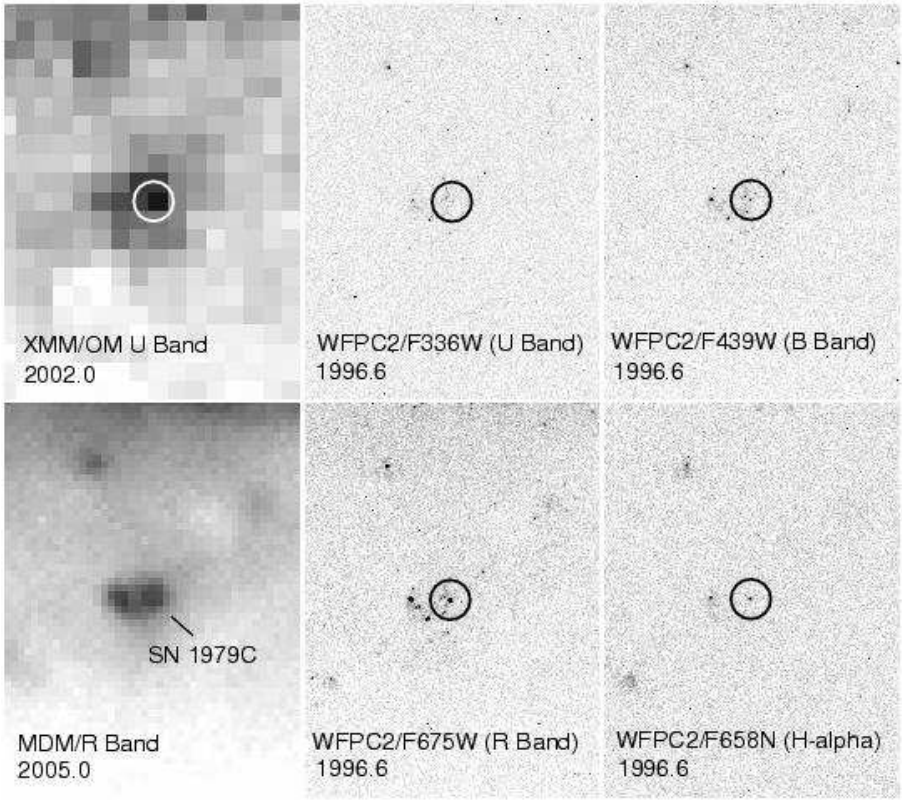,width=12cm,angle=0}
     \end{picture}
\caption{
\X\ OM $U$-band image (from December 28, 2001; upper left-hand panel), 
MDM $R$-band image (from December 19 and 20, 2004; lower left-hand panel), 
and \H\ WFPC2-PC1 $U, B, R$-band and H$\alpha$ images (from July 29, 1996;
middle and right-hand panels) of the \sn\ region. The circles (radius $1''$) 
are at the \H\ $R$-band centroid position of \sn. 
\label{optical}}
\end{figure*}

In order to compare the CSM properties in the vicinity of recent SNe 
for which sufficient data are available, we computed the CSM densities 
as a function of shell expansion radii for 
SN~1994I (Immler et~al.\ 2002), 
SN~1993J (Immler et~al.\ 2001), and
SN~1978K (Schlegel et~al.\ 2004) in an identical manner.
While the CSM density profiles of SNe~1994I, 1993J, and 1978K can
be fitted with a power-law of the form $\rho_{\rm csm} = \rho_0 (r/r_0)^{-s}$
with $s=1.9$ (SN~1994I), $s=1.6$ (SN~1993J), and $s=1.0$ (SN~1978K), we do 
not find a statistically acceptable fit for the CSM density profile using 
all X-ray data of \sn\ (see Table~\ref{tab3} and Fig.~\ref{rho}). 
If we only include the \R\ HRI and \X\ MOS detection (which have the
highest signal-to-noise ratio), we obtain a best-fit CSM density profile 
of $\rho_{\rm csm} = \rho_0 (r/r_0)^{-s}$ with $s=1.6\pm1.2$. 
Inclusion of the early \E\ upper limit gives a profile flatter than 
$s\ls1.7$, consistent with the profile inferred from the \R\ HRI and 
\X\ MOS data. It should be noted, however, that cross-calibration
uncertainties of $\ls10$\% between the \X, \R, and \E\ fluxes
have not been taken into account, which would increase the error to
the CSM profile accordingly (see \S~\ref{results_xmm}).

While the CSM density profile is flat compared to SNe 1994I and 1993J,
it is statistically consistent with a constant mass-loss rate and constant 
wind velocity profile. 
Within the errors, the two data points for the \R\ PSPC and \C\
measurements are consistent with the CSM number density inferred from the 
\R\ HRI and \X\ MOS data, and close to the extrapolated CSM density profile 
of SN~1978K at these large radii from the site of the explosion (log\,$r[{\rm cm}] = 17.50$
and $17.59$; see Fig.~\ref{rho}). A higher assumed wind velocity would 
further decrease the CSM number densities and give an even closer match 
to the extrapolated SN~1978K profile.
Even though clumps in the CSM leading to an increase in the CSM density
cannot be conclusively excluded, the seeming `jump' in the CSM density between 
log\,$r[{\rm cm}] =17.67$ (\R\ PSPC) and $17.82$ (\C\ ACIS data) is likely
caused by the large errors associated with the low photon statistics of
these observations and cross-calibration uncertainties between the
different instruments.

Spatially resolved \VLBI\ radio imaging of \sn\ at an age of $t=22$~yrs
have, in fact, shown a uniform shell-like structure with no evidence
for a clumpy CSM or a geometry different from spherically symmetric
(Bartel \& Bietenholz 2003). Radio monitoring of \sn\ over the past
decades has shown significant quasi-periodic or sinusoidal modulations 
in the radio flux (see Fig.~2 in Montes et~al.\ 2000). These modulations
in the radio lightcurve have been discussed in the context of a modulation
of the stellar wind by a binary companion in a highly eccentric orbit
(Weiler et~al.\ 1992; Montes et~al.\ 2000). 
The lack of a change
in the radio spectral index over the monitored period, however, indicates an 
unchanged emission mechanism with constant efficiency (Montes et~al.\ 2000). 
Our X-ray data do not support a substantial change in the CSM density profile,
but confirm a scenario in which the shock interacts with a dense CSM
($\gs 10^4~{\rm cm}^{-3})$.

\subsection{Optical/UV Data}

Late-time optical and UV photometry and spectroscopy can yield
interesting insights into the CSM properties, such as electron densities,
shock velocities and ionization states of the shocked CSM.
From these, conclusions can be drawn on the nature and mass of
the progenitor, as well as on its evolution.
Of particular interest are the late-time optical
and UV emission as a result of radiative cooling of the SN shock,
ionized by incident X-rays from the CSM shock interaction and by
the SN event itself (Chevalier \& Fransson 1994).

Multi-band \H\ imaging of the environment of \sn\ was successfully used
to recover optical emission from \sn\ at $t \approx 20$~yrs after its
outburst (van Dyk et~al.\ 1999). The \H\ data showed \sn\ to be
the brightest source at any observed wavelength band ($U, B, V, R, I$) within 
a small stellar cluster of red supergiants and young blue stars. 
While \sn\ had magnitudes of 
$m_{\rm U} = 23.2$~mag,
$m_{\rm B} = 23.3$~mag, 
$m_{\rm V} = 22.1$~mag,
$m_{\rm I} = 21.0$~mag,
the next brightest optical/UV sources in the environment had 
$m_{\rm V} \gs 24$~mag and $m_{\rm I} \gs 23$~mag. 
The \H\ images further suggest the presence of 5--7 additional, although 
faint, UV-emitting stars at the site. Three more UV sources were imaged 
with the \H\ further to the east ($\approx2''$), which are located within
the OM counts extraction aperture ($2\farcs9$ radius). 
It is therefore possible that these sources contribute a significant fraction 
to the observed optical/UV flux within the extracted aperture.

The new ground-based $R$-band and H$\alpha$ images taken in December 2004,
some 25.6~yrs after the outburst, continue to indicate a strong CSM interaction,
three years after the \X\ observations. This suggests that both 
H$\alpha$ and the broad [O I] and [O II] line emissions seen in the 1993 
spectra are still of comparable strength to levels more than a decade earlier.

Little is known to date about the late-time spectra of SNe shortward of
5,000~\AA. UV emission is detected from the position of \sn\ with a 
$UVW1$-band luminosity of $L_{\rm UVW1}=9 \times 10^{36}~{\rm ergs~s}^{-1}$.
Comparison of the H$\alpha$ and UV luminosities of \sn\ show 
an interesting similarity 
($L_{\rm H\alpha} = 1.6 \times 10^{37}~{\rm ergs~s}^{-1}$,
van Dyk et~al.\ 1999). The H$\alpha$ lightcurve of 
\sn\ shows no decline between 1987 and 1991 (Fesen et~al.\ 1999),
while a slight increase in the 1993 data is observed, which 
coincides with the observed flattening of the radio lightcurve. 
A similar H$\alpha$ luminosity was observed
for the Type II-L SN~1980K ($2.5 \times 10^{37}~{\rm ergs~s}^{-1}$;
Fesen et~al.\ 1999) at an age of 14--15 years after its explosion.
These high late-time H$\alpha$ luminosities are largely the result 
of high electron densities ($n_{\rm e} \gs 10^4$--$10^5~{\rm cm}^{-3}$),
which are supported by the high inferred CSM densities in this paper
($\gs 10^4~{\rm cm}^{-3}$).

The OM $U$-band data show a bright source at the position of \sn\
($L_{\rm U} = 5 \times 10^{36}~{\rm ergs~s}^{-1}$), while the \H\ 
data from 1996 show very weak emission from \sn\ in the $U$-band.
This difference can be attributed to the cut-off of the \H\ $U$-band
filter redward to 3,800~\AA, while the $U$-band filter of the OM
cuts off at 3,950~\AA, which then includes the broad [Ne III] emission
lines at $\lambda\lambda$3868,3869. The detection of [Ne III] emission lines from
\sn\ (Fesen et al.~1999) may explain why the SN is brighter in the \X\ OM 
$U$-band images than in the \H\ $U$-band (F336W filter) images.

The peak surface brightness of the OM $U$-band emission coincides with 
the peak in the $R$-band (see Fig.~\ref{optical}), which includes broad 
H$\alpha$, [O II] $\lambda\lambda$7319,7330, and [O I] 
$\lambda\lambda$6300,6364 line emission from the SN ejecta, 
and narrow H$\alpha$, presumably associated with the strong CSM interaction.
While the source at the position of \sn\ is classified as point-like
in all OM bands by the source detection routines, slightly more extended 
emission is visible in the OM $B$ and $UVW1$-band images compared to the
$U$-band. This is consistent with at least a substantial fraction of the 
$B, U$ ($\ls30\%$) and $UVW1$-band ($\gs50\%$) fluxes being to the SN itself.
For the $UVW1$-band, most of the flux likely arises from Mg II 2,800~\AA\ 
where the peak in the $UVW1$-band effective area is located.
A substantial fraction of the $B$-band flux might also be due to \sn\
(20--30\% contribution to the total light in this OB/SN spatial region) 
and is due to [O III] 4,363~\AA\ line emission.
The contribution from the OB stars to the east may be significantly 
reduced, relative to that from the SN, to the total flux within the 
$U$-band passband for both the \H\ and OM filters, due to the Balmer
jump at 3,600~\AA\ in the stellar spectra of these stars.

Because of the strong appearance of \sn\ in the OM $U$-band image, the 
[Ne III] emission may have substantially increased in strength since the 
\H\ spectra obtained in 1996 (Fesen et al.~1999). It should be kept in mind, 
however, that the 
\H\ data were acquired before the increase of \sn\ at radio wavelengths.
This increase could be caused by either a higher ionization of the CSM 
interaction and/or a lowering of the local reddening/extinction due to dust 
evaporation via shocks (i.e., dust in the CSM) or less ejecta formed dust
via dilution from expansion.

In summary, the high inferred optical/UV flux for \sn, especially in
the $U$-band, further suggests continued strong CSM interaction even at this 
late stage in the evolution and large radii from the site of the explosion,
which is also supported by our recent MDM $R$-band data,
and confirms our results which are independently inferred from the X-ray data.

\section{Summary}
\label{summary}

In this paper we present a comprehensive study of the Type II-L
\sn\ in X-rays and in the optical/UV based on \X\ data, all archival X-ray 
data available to date, as well as archival optical data from the \H\ and 
a recent MDM observation. Key findings are: \\

\noindent 
$\bullet$~\sn\ is observed at a high X-ray luminosity 23~yrs after its 
outburst ($L_{0.3-2}=8\times10^{38}~{\rm ergs~s}^{-1}$) and shows no 
evidence for a decline over the observed period of $16$--$23$~yrs after 
its outburst.

\noindent 
$\bullet$~The X-ray spectrum, measured for the first time, gives a 
best-fit two-temperature thermal plasma model with 
$kT_{\rm high}=4.1^{+76}_{-2.3}$~keV, 
$kT_{\rm low}=0.76^{+0.25}_{-0.17}$~keV and Galactic absorption only,
characteristic for emission from both the forward (hard component)
and reverse shock (soft component). The spectral properties are similar 
to that of other core-collapse SNe, such as SNe 1978K and 1993J.

\noindent 
$\bullet$~After correction for a decelerating shock front we find no
evidence for a change in the progenitor's mass-loss rate 
($\dot{M} \approx 1.5 \times 10^{-4}~M_{\odot}~{\rm yr}^{-1}$) 
over a period of $\gs16,000$~yrs in the stellar evolution of the progenitor.

\noindent 
$\bullet$~High CSM number densities ($\gs 10^4~{\rm cm}^{-3}$) are inferred 
for \sn\ even at large radii (out to $\approx 4 \times 10^{17}~{\rm cm}$) 
from the site of the explosion, similar to SN~1978K.
The CSM density profile can be fitted with a power law of
the form $\rho_{\rm csm} = \rho_0 (r/r_0)^{-s}$ with index $s \ls 1.7$
comparable to other SNe (e.g., SNe 1994I and 1993J), and support a constant 
mass-loss rate and a constant stellar wind velocity.

\noindent 
$\bullet$~A point-like optical/UV source is detected at the position
of \sn\ with the Optical Monitor on-board \X, with
$U$, $B$, and $UVW1$-band luminosities of 
$L_{\rm U} = 5 \times 10^{36}~{\rm ergs~s}^{-1}$,
$L_{\rm B} = 7 \times 10^{36}~{\rm ergs~s}^{-1}$, and 
$L_{\rm UVW1} = 9 \times 10^{36}~{\rm ergs~s}^{-1}$,
similar to its H$\alpha$ and $R$-band luminosity.
The high luminosities, especially in the $U$-band arising from
Mg II 2,800~\AA\ and in the $UVW1$-band from [Ne III]
$\lambda\lambda$3868,3869 emission lines, indicate substantial CSM interaction
in the late-time evolution of \sn\ and independently support
the strong observed CSM interaction as observed in X-rays.

\acknowledgments

This paper is based on observations obtained with XMM-Newton, an ESA 
science mission with instruments and contributions directly funded by 
ESA Member States and NASA.
The research has made use of data obtained through the High Energy 
Astrophysics Science Archive Research Center Online Service, provided by 
the NASA/Goddard Space Flight Center.
KWW wishes to thank the Office of Naval Research (ONR) for the 6.1 
funding supporting his research.


\newpage
\begin{deluxetable}{cccccc}
\tabletypesize{\footnotesize}
\tablecaption{X-Ray Observations of SN~1979C \label{tab1}}
\tablewidth{0pt}
\tablehead{
\colhead{Mission} &
\colhead{Instrument} &
\colhead{Obs-ID/Seq-No} &
\colhead{Date} &
\colhead{Exposure} &
\colhead{MJD}
}
\startdata
\E\ & HRI & H1220N16.XIA & 1979-06-07 & \phantom{0}4.0~ks  & 44,031  \\ 
\E\ & HRI & H1220N16.XIB & 1979-12-12 & 13.4~ks  & 44,217  \\ 
\E\ & HRI & H1220N16.XIC & 1980-06-29 & 23.8~ks  & 44,418  \\ 
\R\ & HRI & RH600731N00 & 1995-06-17 & 42.8~ks & 49,885  \\
\R\ & HRI & RH500422N00 & 1995-07-01 & \phantom{0}9.0~ks & 49,899  \\
\A\ & SIS & 55044000 &	  1997-12-24 & 27.3~ks & 50,806  \\
\R\ & HRI & RH500542N00 & 1997-12-27 & 25.1~ks & 50,809  \\
\R\ & PSPC & RP180296N00 & 1998-12-08 & \phantom{0}7.8~ks & 51,155 \\ 
\C\ & ACIS-S & 400  &	1999-06-11 & \phantom{0}2.5~ks & 51,340  \\
\X\ & MOS & 0106860201 & 2001-12-28 & 36.6~ks &	52,271
\enddata
\tablecomments{
All \E\ HRI observations were merged into a single 41.2~ks observation.
}
\end{deluxetable}


\newpage
\begin{deluxetable}{cccccc}
\tabletypesize{\footnotesize}
\tablecaption{Optical/UV Observations of SN~1979C \label{tab2}}
\tablewidth{0pt}
\tablehead{
\colhead{Telescope} &
\colhead{Filter} &
\colhead{Date} &
\colhead{Exposure} &
\colhead{MJD}
}
\startdata
\H\ WFPC2-PC1 	 & U (F336W) & 1996-07-29 & 2.6~ks & 50,293  \\
\H\ WFPC2-PC1 	 & B (F439W) & 1996-07-29 & 2.4~ks & 50,293  \\
\H\ WFPC2-PC1 	 & R (F675W) & 1996-07-29 & 1.0~ks & 50,293  \\
\H\ WFPC2-PC1 	 & H$\alpha$ (F658W) & 1996-07-29 & 3.9~ks & 50,293  \\
\X\ OM   	 & U &  2001-12-28 & 1.8~ks & 52,271  \\
\X\ OM 		 & B &  2001-12-28 & 2.0~ks & 52,271  \\
\X\ OM 		 & UVW1 & 2001-12-28 & 1.9~ks & 52,271  \\
MDM Hiltner 2.4m & R  & 2004-12-19 & 2.0~ks & 53,358  \\
MDM Hiltner 2.4m & H$\alpha$  & 2004-12-19 & 3.0~ks & 53,358
\enddata
\end{deluxetable}


\begin{deluxetable}{cccccccc}
\tabletypesize{\footnotesize}
\tablecaption{X-Ray Properties of SN~1979C \label{tab3}}
\tablewidth{0pt}
\tablehead{
\colhead{Epoch} &
\colhead{Instrument} &
\colhead{$f_{0.3-2}$} &
\colhead{$L_{0.3-2}$} &
\colhead{$\dot{M}$} &
\colhead{log\,$r$} &
\colhead{$t_{\rm w}$} &
\colhead{$n$} \\
${\rm [yrs]}$ & & [$10^{-14}~{\rm ergs~cm}^{-2}~{\rm s}^{-1}$] & [$10^{39}~{\rm ergs~s}^{-1}$]
 & [$10^{-4}~M_{\odot}~{\rm yr}^{-1}$] & [cm] & [yrs] & [$10^4~{\rm cm}^{-3}$] \\
\noalign{\smallskip}
(1) & (2) & (3) & (4) & (5) & (6) & (7) & (8)
}
\startdata
\phantom{0}0.66 & \E\ HRI &  $<12.78$ &	$<4.47$ & 	$<0.71$ &	16.28
     & 	\phantom{0}\phantom{0}\,615 & $<223.87$ \\
16.18 & \R\ HRI &  $2.20\pm0.37$ & $0.77\pm0.13$ &	$1.45\pm0.25$ &	17.50 	
     & 	14,878 & $1.17\pm0.32$ \\
16.22 & \R\ HRI &  $<5.09$ & 	$<1.78$	 & 	$<2.19$  &	17.51 
     & 	14,914 & $<1.17$ \\
18.71 & \A\ SIS &  $<4.92$ & 	$<1.72$ & 	$<2.33$ &	17.51 
     & 	17,200 & $<1.66$ \\
18.72 & \R\ HRI &  $<5.15$ & 	$<1.80$ & 	$<2.36$ &	17.51	
     & 17,208	& $<1.66$ \\
19.67 & \R\ PSPC & $2.21\pm0.63$ &	$0.77\pm0.22$ & $1.62\pm0.46$ &	17.52
     & 	18,080 & $1.66\pm0.63$ \\
20.17 & \C\ ACIS-S&$2.23\pm0.71$ &	$0.78\pm0.25$ &	$1.63\pm0.52$ &	17.53 	
     & 18,546 	& $1.51\pm0.56$ \\
22.72 & \X\ MOS &  $2.31\pm0.26$ &	$0.81\pm0.09$ &	$1.75\pm0.19$ &	17.59	
     & 20,893 & $0.85\pm0.19$
\enddata
\tablecomments{
(1)~Epoch after the peak optical brightness (April 15, 1979) in units of yrs;
(3)~0.3--2~keV X-ray band flux in units of ${\rm ergs~cm}^{-2}~{\rm s}^{-1}$;
(4)~0.3--2~keV X-ray band luminosity in units of ${\rm ergs~s}^{-1}$;
(5)~mass-loss rate of the progenitor in units of $M_{\odot}~{\rm yr}^{-1}$;
(6)~log of the radius from the site of the explosion;
(7)~age of the stellar wind in units of yrs;
(8)~post-shock CSM number density in units of $10^4~{\rm cm}^{-3}$.
}
\end{deluxetable}
\vfill


\clearpage
\vfill

\vfill

\vfill

\vfill

\vfill

\vfill

\vfill

\end{document}